# Biases in differential expression analysis of RNA-seq data: A matter of replicate type

Sora Yoon[1], Dougu Nam[1,2,*]

[1]School of Life Science, Ulsan National Institute of Science and Technology, Ulsan, Republic of Korea

[2]Department of Mathematical Sciences, Ulsan National Institute of Science and Technology, Ulsan, Republic of Korea

* To whom correspondence should be addressed. Tel: +82-52-217-2525; Fax:+82-52-217-2639;

 Email: dougnam@unist.ac.kr

**ABSTRACT**

In differential expression (DE) analysis of RNA-seq count data, it is known that genes with a larger read number are more likely to be differentially expressed. This bias has a profound effect on the subsequent Gene Ontology (GO) analysis by perturbing the ranks of gene-sets. Another known bias is that the commonly used parametric DE analysis methods (e.g., edgeR, DESeq and baySeq) tend to yield more DE genes as the sequencing depth is increased. We nevertheless show that these biases are in fact confined to data of the technical replicate type. We also show the GO or gene-set enrichment analysis methods applied to technical replicate data result in considerable number of false positives. In conclusion, the current DE and enrichment analysis methods can be confidently used for biological replicate count data, while caution should be exercised when analysing technical replicate data.

**INTRODUCTION**

High-throughput RNA sequencing (RNA-seq) provides portraits of the transcriptome landscape at an unprecedented resolution [1,2]. RNA-seq typically produces millions of sequencing reads, each of which provides a bit of information for genomic events in the cell. Thus, unlike microarray, RNA-seq has highly flexible applications for diverse genomic analyses such as quantitation of gene expression, finding of new transcripts, detection of single nucleotide polymorphisms, RNA editing, gene fusion detection and so on [3-8]. Among these applications, the quantitation of gene expression may be one of the basic functions of RNA-seq. It is performed by simply counting the reads aligned to each gene or exon region. RNA-seq also has advantages in this application over microarrays in both reproducibility and detection of weakly expressed transcripts [9].

Molecular biology research is focussed on questions such as 'what happens in the cell' and 'what changes between differing cell conditions'. While the sequencing technology has shown advantages for answering the former question, it also gave rise to some complicated issues with the latter, for the following aspects. (1) *normalization*: In contrasting RNA-seq counts between cell conditions, each sample can have different sequencing depths and RNA compositions. Therefore, appropriate normalization should be applied to make the gene expression levels comparable or to estimate the model parameters [10-12]. (2) *probability modelling*: Since this is counting data, discrete probability models (Poisson or negative binomial model) have been used to assess the differential expression (DE) of genes. Parameter estimation is a critical issue especially for small sample sizes [9,13,14]. (3) *biases in DE analysis*: a striking bias with DE analysis of RNA-seq count data was found in that genes with a larger read number (or longer genes) have a greater likelihood of being detected, which we may call the *read number bias* [15]. This bias hampered the subsequent Gene Ontology (GO) category enrichment analysis such that GO terms annotated to many long genes had a greater chance of being selected. A resampling based method was eventually developed to account for such selection bias in GO analysis [16]. Despite the profound effect that the read number bias might have on DE and subsequent analyses, this bias has often been ignored resulting in confusion. Another known bias is that commonly used parametric DE analysis methods edgeR [14], DEseq [13] and baySeq [17], tend to yield more DE genes as the sequencing depth is increased. In contrast, a nonparametric method was suggested to have consistent DE calls that are independent of the sequencing depth [18].

In this study, it is shown that these biases regarding DE analysis of RNA-seq data are confined to the technical replicate data obtained from the same samples. In particular, the relatively *small variances of genes* in the technical replicate data cause these biases. It is also shown that the GO category and gene-set enrichment analysis (GSEA) methods [19,20] applied to technical replicate data result in a considerable number of false positives. This indicates the read number bias is another source of false positives in enrichment testing besides the well-known 'inter-gene correlations' [20,21]. On the other hand, DE analysis with the biological replicate data obtained from different samples exhibited no such biases, and parametric DE and subsequent enrichment tests can be used without any special concern regarding those biases and the false positives.

**RESULTS**

**The read number bias is observed only for technical replicate type data**

In DE analysis of RNA-seq count data, it is known that genes with larger read numbers (or longer genes) are more likely to be differentially expressed [15,16]. We examined if there exists such a pattern by plotting the gene differential scores (SNR: signal to noise) for four RNA-seq read count datasets, with each having two sample groups. The four datasets used in this study are denoted as Marioni, MAQC-2, TCGA KIRC and TCGA BRCA, respectively. See the Methods section for the description of the datasets. The 'read number bias' was well represented in the first two datasets (Marioni and MAQC-2) as shown in Figure 1, where genes with a larger read number had more dispersed distributions of gene scores. This pattern indicates that genes with larger read numbers are more likely to have a higher level of differential scores. However, many of the read count data from TCGA (The Cancer Genome Atlas) [22] did not show such a bias but exhibited an even dispersion as illustrated in the lower figures in Figure 1. A key factor between the two distinctly different dispersion patterns was the sample replicate type: The former two (Marioni and MAQC-2 dataset) were composed of technical replicate samples while the latter two (TCGA KIRC and TCGA BRCA) of biological replicates. Furthermore, we selected five available DE analysis methods for RNA-seq count data to examine how DE genes are distributed across different read numbers in each method. Among the five DE analysis methods, three (edgeR, DESeq, baySeq) are parametric methods while the other two (NOISeq, SAMseq) are nonparametric. For all of the five methods, genes with a larger read number had a higher proportion of DE genes for both technical replicate datasets, while such bias was not seen for either of the biological replicate datasets (Supplemental Fig. 1). This indicates the read number bias is a characteristic of the sample replicate type rather than the DE analysis methods or the counting data type. To corroborate the evidence, we analysed probability models and conducted a simulation study, as described in the next section.

**Gene-wise variance in the read count data determines the read number bias**

The technical replicate data are generated from the same samples, so most of its variation comes from experimental noise. In such a case, $X_{ij}$, the read count of $i$th gene in $j$th sample can be simply assumed to have a Poisson distribution $X_{ij} \sim Poisson(\mu_{ij})$ the mean and variance of which are the same as $\mu_{ij}$ [9]. However, for biological replicates, additional variations between individuals are involved [13,23]. In such a case, the read count $X_{ij}$ is modelled by a negative binomial (NB) distribution to account for the increased variation, and denoted as $X_{ij} \sim NB(\mu_{ij}, \sigma_{ij}^2)$ where $\mu_{ij}$ and $\sigma_{ij}^2$ are the mean and variance, respectively. Its variance is given as $\sigma_{ij}^2 = \mu_{ij} + \alpha_i \mu_{ij}^2$, which is not less than the mean $\mu_{ij}$ and $\alpha_i \geq 0$ is the dispersion coefficient for gene $i$, which determines the amount of additional variability. In particular, the NB distribution degenerates to a Poisson distribution when $\alpha_i$ approaches 0. Thus, the SNR score for the biological replicate data is represented as

$$SNR_i = \frac{\mu_{i1} - \mu_{i2}}{\sigma_{i1} + \sigma_{i2}} = \frac{\mu_{i1} - \mu_{i2}}{\sqrt{\mu_{i1} + \alpha_i \mu_{i1}^2} + \sqrt{\mu_{i2} + \alpha_i \mu_{i2}^2}}, \qquad (1)$$

where $\mu_{ik}$ is the arithmetic mean of the counts for $i$th gene in the sample group $k$=1,2.

For the technical replicate case, the dispersion coefficient $\alpha_i$ is close to 0, and the SNR value is

$$SNR_i \approx \frac{\mu_{i1} - \mu_{i2}}{\sqrt{\mu_{i1}} + \sqrt{\mu_{i2}}} = \sqrt{\mu_{i1}} - \sqrt{\mu_{i2}}$$

which directly depends on the read number. This accounts for the increasing SNR variation in the upper figures in Figure 1. However, for biological replicate data, $\alpha_i$ is not negligible in (1) and the SNR is estimated as follows: Without a loss of generality, suppose $\mu_{i1} > \mu_{i2}$, then $|SNR_i| \leq \frac{|\mu_{i1} - \mu_{i2}|}{\sqrt{\mu_{i1} + \alpha_i \mu_{i1}^2}} \to \frac{1}{\sqrt{\alpha_i}}$ as $\mu_{i1}$ increases. Thus, the SNR has bounded values irrespective of the read numbers.

Based on this rationale, we simulated read count data to test how the SNR scores are distributed for each replicate model. The technical and biological replicate data were generated using Poisson and negative binomial distributions, respectively; 30% of the genes were chosen and their test group counts were 1.3-fold increased or decreased to generate DE genes (see Methods). Then, the SNR values for each replicate type were depicted as shown in Figure 2, which exactly reproduced the SNR patterns observed with the real count datasets. With the simulated technical replicate data, the SNR scores of the DE genes (red dots) became more dispersed as their read numbers increased (Fig. 2a). On the other hand, with the simulated biological replicates, the SNR scores of the DE genes were deemed independent of the read numbers (Fig. 2b). This indicates the Poisson like variation in the technical replicate data is the primary cause of the read number bias in the DE analysis of read count data.

**Sequencing depth bias originates from the small variance of genes in the technical replicate data.** There is a related problem regarding sequencing depth. Tarazona and colleagues proposed a nonparametric method (NOISeq) for DE analysis of RNA-seq [18]. They found that parametric DE analysis methods (edgeR, DESeq, baySeq) tend to yield more DE genes as the sequencing depth is increased, while their nonparametric method yields an even number of DE genes regardless of depth. It was also demonstrated that a substantial portion of the increased DE genes obtained with parametric methods are false positives, based on the assumption that genes with a p-value<0.001 and two or larger fold changes in the qPCR data are 'gold standard' true positives. Because the two-fold criterion may not be generally considered in DE analysis, we focus on the DE number bias against the sequencing depths.

Since only technical replicate datasets were used for analysing the sequencing depth bias, we investigated whether such bias with parametric methods is still found with biological replicate data. Read count data at different sequencing depths were simulated by using the quartiles of the read counts (i.e. 0.25, 0.5 and 0.75 were multiplied to the count data and rounded to integer values). First, the three parametric methods edgeR, DESeq and baySeq were applied to the two technical replicate datasets (Marioni, MAQC-2), and the number of DE genes for different sequencing depths were depicted. Then, the same test was conducted for the two biological replicate datasets (TCGA KIRC, TCGA BRCA) (Fig. 3). To our surprise, the depth bias of the parametric methods was observed only with the technical replicate datasets. This indicates that some characteristic of the technical replicate data may have caused the depth bias in the parametric methods. To pinpoint the cause of the depth bias, the simulation datasets generated by Poisson and NB distributions were used again, and the depth bias was only observed with the Poisson distributed dataset (Supplemental Fig. 2). This indicates the small variance of genes in the technical replicate data rather than the DE analysis methods is the cause of the sequencing depth bias.

**Small variance in technical replicate data results in false positives in gene set testing methods**

Lastly, the effect of the replicate type on gene-set testing methods was analysed. We show that the small variance of genes in the technical replicate data remarkably increases false positives in GO analysis [19] and GSEA [20]. To measure the amount of increased false positives in gene-set tests, the false enrichment rate (FER) was devised as follows:

$$\text{FER} = \frac{Average\ \#\ of\ enriched\ gene\ sets\ in\ gene-permuted\ datasets}{\#\ of\ enriched\ gene\ sets\ in\ original\ dataset}$$

That is, FER is the ratio of the significant gene sets in gene-permuted datasets to those of its original dataset with the same FDR cutoff applied to both numerator and denominator. Note that most of the significant (=enriched) gene-sets in gene-set testing methods are expected to become non-significant if the gene labels are permuted.

The FER values were compared in GO analysis and GSEA for both technical and biological replicate datasets (Marioni, MAQC-2, TCGA KIRC and TCGA BRCA), and the MSigDB C5 (GO) gene sets [24] were used for the tests.

For GO analysis, the lists of DE genes were determined using DESeq method for a number of different cutoffs, and the phyper R function was used to assess the enrichment of each GO set in each of the DE gene list. For each dataset, gene permutation was performed ten times and the average number of significant gene-sets was used to calculate FER. Figure 4 shows clearly different patterns of FERs between the technical and biological replicate datasets. The FER values for the biological replicate datasets were nearly zero, indicating no falsely enriched gene sets. In contrast, very high FER values were observed for the technical replicate datasets, which indicates a biased DE distribution (upper figures in Fig. 1) by itself generates a great number of falsely enriched gene-sets. Thus, the read number bias with the GO analysis of technical replicate data not only perturbs the ranks of gene sets, it also considerably increases false positives. A similar tendency was observed with the GSEA, which was implemented using FDR cutoff 0.25 and GSEA-R [20]. The FERs of the biological replicate datasets were only 0.029 (0.5/17, TCGA KIRC) and 0.023 (1.5/66, TCGA BRCA), while those of the technical replicate datasets reached as high as 0.894 (679.5/760, Marioni) and 0.927 (626/675, MAQC-2), respectively. Such high FERs for the technical replicate data indicate the read number bias is another source of false positives besides the inter-gene correlation, and these false positives cannot be removed even by the sample permutation procedure of GSEA. Therefore, great care should be taken when applying GO analysis or GSEA to technical replicate data, while they can be applied to biological replicate data with confidence.

**Discussion**

Previous studies have reported biases in differential analysis of RNA-seq count data regarding gene length (read number), sequencing depth and GO analysis [15,16,18]. However, to the best of our knowledge, all of these were derived based on technical replicate data, which may not be generalizable to biological replicate data.

Indeed, it was demonstrated in this study that such biases are confined to only the technical replicate, so the current DE and gene-set analysis methods can be safely applied to biological replicate count data without any involvement of these biases. To this end, mathematical inferencing, model-based simulation and tests with real data were performed to pinpoint the small variance of genes in the technical replicate data as the common cause of these biases as well as the false positives that occur in gene-set analysis methods. Thus, researchers should be very cautious when applying DE and gene-set analysis to RNA-seq data of technical replicate type.

As the cost of sequencing continues to decrease, more and more biological replicate data will come to be used in near future. The analysis presented here provides a unified perspective that can explain the known biases in DE analysis and the false positives in gene-set analysis of RNA-seq data. This may help remove unnecessary concern and confusion in RNA-seq data analysis.

**Methods**

**Datasets**

We used four publicly available RNA-seq read count datasets including two technical replicate type (Marioni and MAQC-2) and two biological replicate type (TCGA KIRC and TCGA BRCA) datasets.

- Marioni dataset: Total mRNAs extracted from human liver and kidney sample were sequenced on 14 lanes of Illumina GA 1 (seven lanes each). In the early stage of RNA-seq study, Marioni and colleagues produced this dataset to estimate the technical variance of Illumina sequencing and compared the capability of identifying DE genes with the Affymetrix array. A processed RNA-seq read count matrix was downloaded from the supplemental table 2 of the Marioni and colleagues' paper [9].

- MAQC-2 dataset: The MicroArray Quality Control Project (MAQC) measured the gene expression levels from two distinct RNA samples (Ambion's human brain reference RNA and Stratagene's human universal reference RNA) on four titration pools on seven microarray platforms [25]. MAQC-2 is a part of MAQC project. Bullard and colleagues [10] sequenced the two RNA samples on Illumina Genome Analyser II and then, evaluated the statistical methods for normalization and differential expression. The MAQC-2 count data and phenotype information were downloaded from the ReCount [26] web page.

- TCGA KIRC vs normal dataset: The RNA-seq count data of renal clear cell carcinoma (KIRC) patients' tumour and matched normal samples were obtained from The Cancer Genome Atlas (TCGA) data portal [22,27].

- TCGA BRCA dataset: The RNA-seq count data of the invasive breast cancer patients' tumour and matched normal samples were downloaded from the TCGA Data portal.

For TCGA count data, tumour and matched normal samples of randomly selected 7 patients were used for analysis to compare with the technical replicate datasets of similar sample sizes. Also, the genes with 5 or lower median read counts was filtered out in each dataset. The FDR cutoffs for DE, $10^{-4}$ and 0.05 were used for the technical and biological replicate datasets, respectively, because Marioni or MAQC-2 dataset were composed of different tissue types yielding a great number of DE genes.

**Simulation of read count data**

The read count $X_{ij}$ of gene $i$ and sample $j$ was generated using Poisson or negative binomial distribution depending on the sample replicate type

$$X_{ij} \sim Poisson(\mu_{ij}) \quad \text{for technical replicate type}$$

$$X_{ij} \sim NB(\mu_{ij}, \phi_{ij}) \quad \text{for biological replicate type}$$

where $\mu_{ij}$ is the mean and $\phi_{ij}$ is the dispersion parameter. Each simulation data matrix contained 2000 genes and 20 samples (10 samples for each group). The mean read counts for simulated genes were determined by randomly selecting 2000 median gene counts from the TCGA KIRC normal samples. To generate DE genes, 1.3 was either multiplied or divided to the gene's mean for 600 randomly chosen genes (30% of the simulated genes). The dispersion parameters in each gene were also estimated from TCGA KIRC normal data using edgeR package [14] and applied for the simulation. Then, using rpois and rbninom functions in R stats package, technical and biological replicate data were simulated, respectively.

## Acknowledgements

This work was supported by Basic Science Research Program through a National Research Foundation (NRF) grants funded by the Korean government (MSIP & MOE) (2014M3C9A3068555 and 2014R1A1A2056353)

**Author Contributions**

SY and DN conceived the problem and wrote the manuscript. SY analyzed data and DN designed the overall study. All authors reviewed the manuscript.

**Additional information**

**Supplementary information** accompanies this paper at http://www.nature.com/
Scientificreports
**Competing financial interests**: The authors declare no competing financial interests.

# Figures

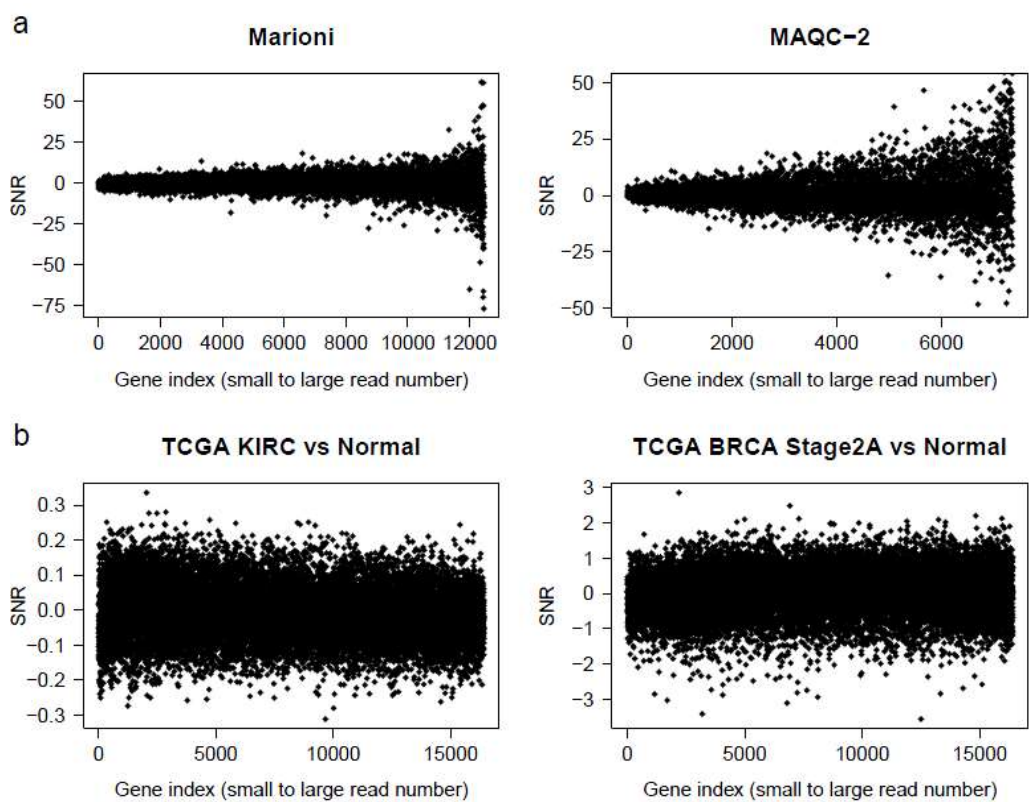

**Figure 1. Signal-to-noise ratio (SNR) distribution in RNA-seq count data.** SNR values were plotted against the read numbers. (a) Technical replicate type data (Marioni and MAQC-2 dataset) exhibited read number bias while (b) biological replicate type data (TCGA KIRC and BRCA dataset) were free from such bias.

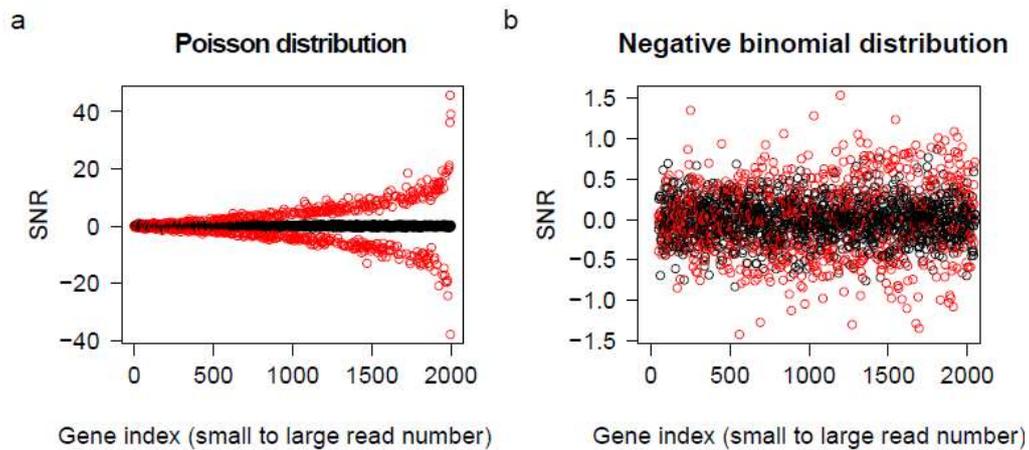

**Figure 2. Signal-to-noise ratio (SNR) distribution in simulated RNA-seq count data.** Read count data were generated using Poisson and negative binomial distributions to simulate (a) technical and (b) biological replicate data, respectively. The SNR values were plotted for each dataset. 600 out of 2000 genes were set as DE genes (red dots) and the others as non-DE genes (black dots).

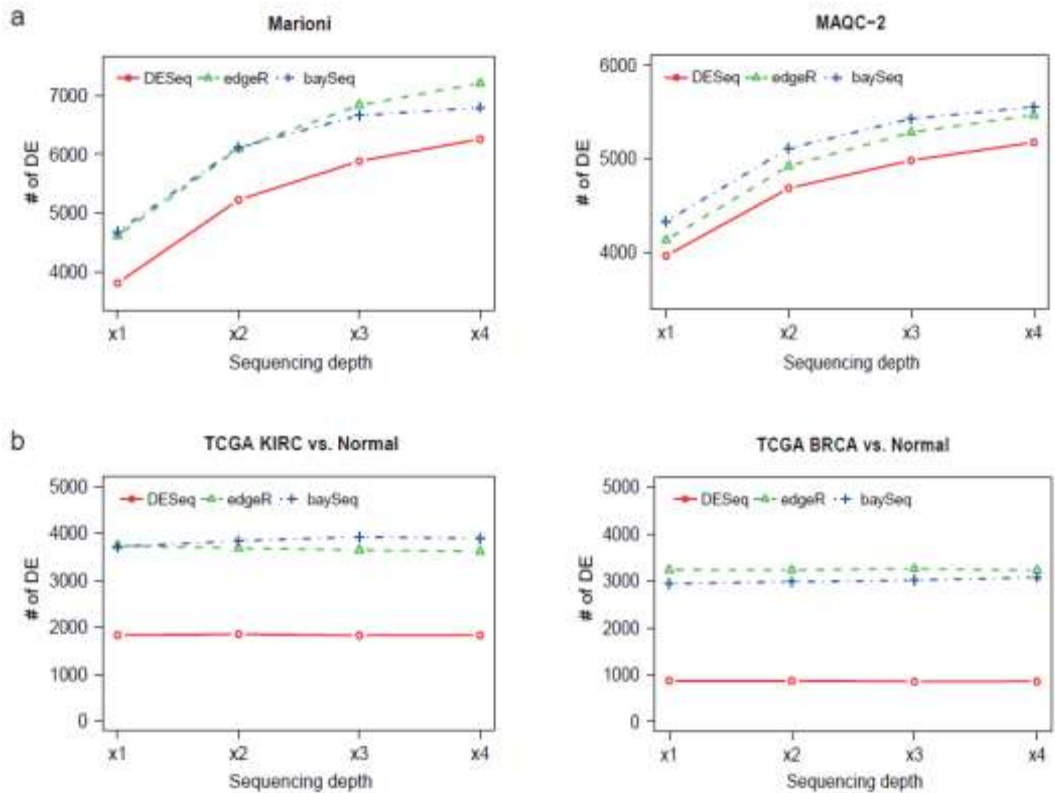

**Figure 3. The number of DE genes against the sequencing depth.** Three parametric DE analysis methods DESeq (red circle), edgeR (green triangle) and baySeq (blue cross) were applied to the four read count datasets (Marioni, MAQC-2, TCGA KIRC, TCGA BRCA) of four different sequencing depths. Data with different sequencing depths were simply obtained using the quartiles of the read count numbers. x1, x2, x3 and x4 represent 1/4, 2/4, 3/4 and 4/4-fold of the original read numbers.

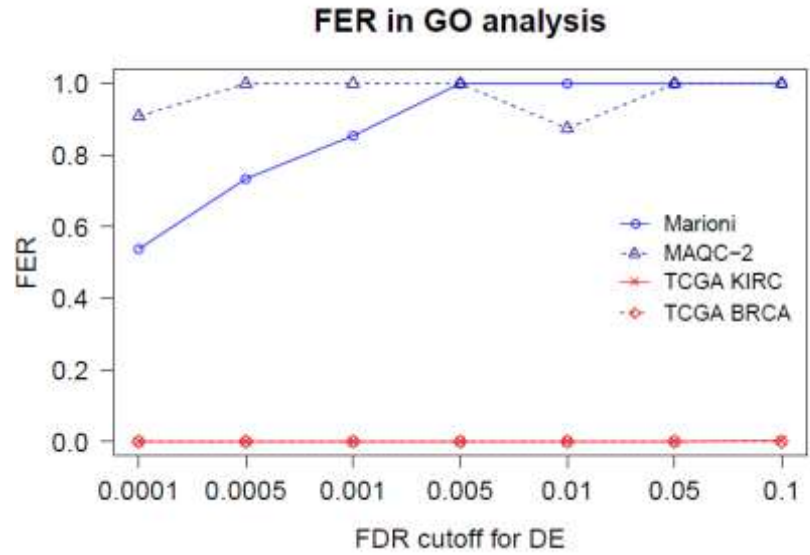

**Figure 4. False enrichment rate in GO analysis.** The false enrichment rate (FER) of gene sets in GO analysis for the two technical (Marioni: blue circle, MAQC-2: dark blue triangle) and two biological replicate (TCGA KIRC: red x, TCGA BRCA: dark red diamond) datasets were plotted against different DE cutoffs for. FERs larger than 1 were truncated to 1.

# Supplementary Figures

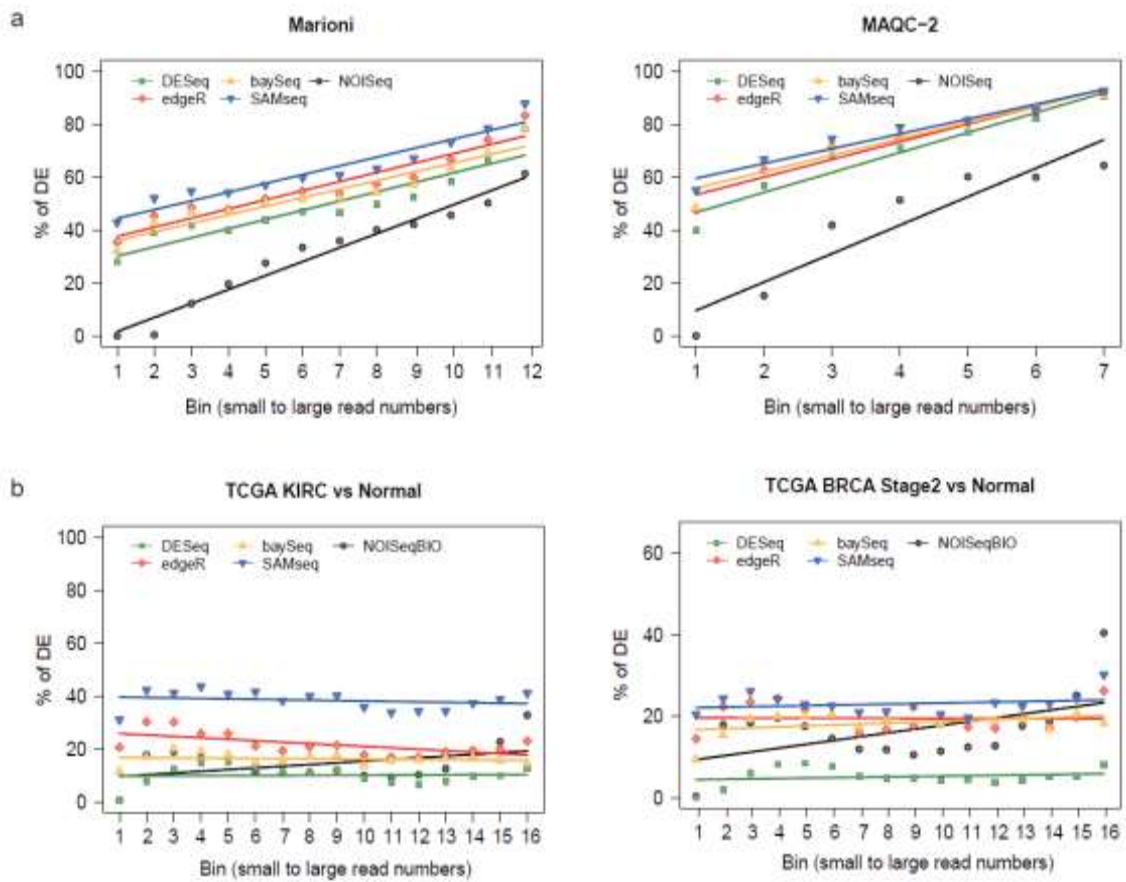

**Supplemental figure 1**. **The proportion of DE genes against the read numbers**. Genes were binned according to the read numbers (bin size: 1000) and then the proportion of DE genes in each bin was calculated. Thus, these graphs represent the overall trend of the DE gene proportion against the read numbers. (a) In both the technical replicate datasets, the number of DE genes kept increasing as the read number was increased no matter which DE analysis method was applied. On the other hand, (b) in the two biological replicate datasets, the number of DE genes were largely independent of the read numbers in most of the DE analysis methods. FDR cutoffs for technical and biological replicates were $10^{-4}$ and 0.05, respectively except for the NOISeq and NOISeqBIO. NOISeq and NOISeqBIO use a joint probability (P) of fold change and mean difference instead of p-value for determining DE genes. We used P=0.9 as a criterion of DE. Among them, NOISeqBIO was recommended for biological replicate data in place of NOISeq.

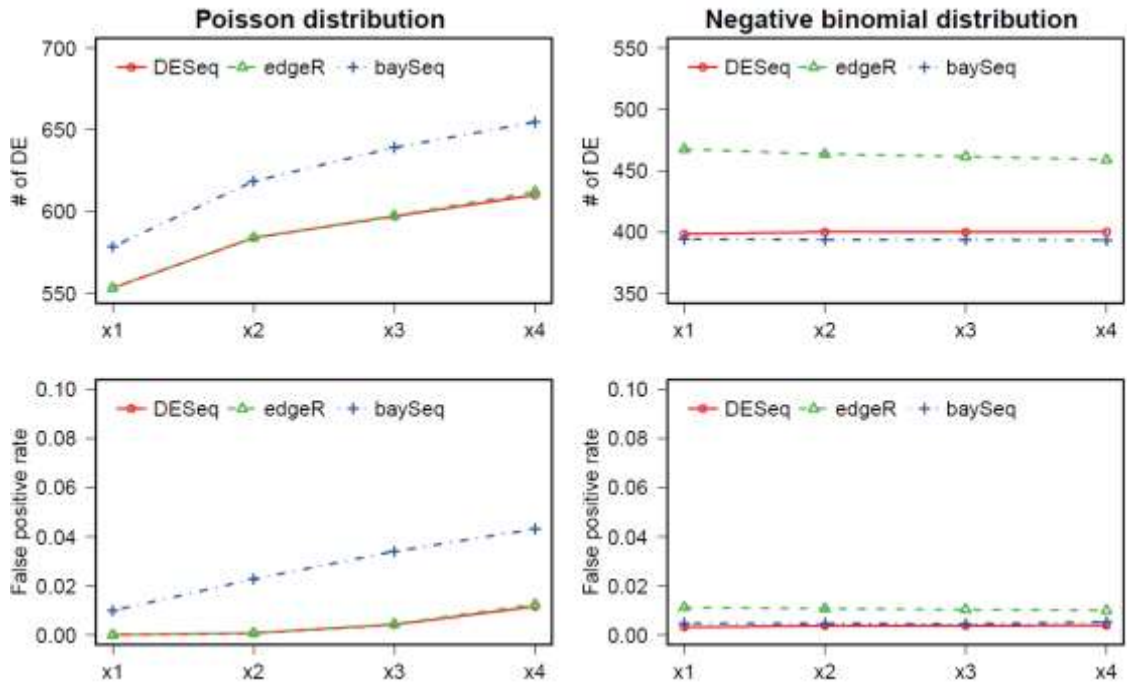

**Supplemental figure 2. Testing sequencing depth bias in simulation datasets.** The number of DE identified by three parametric DE analysis methods (DESeq, edgeR, baySeq) and false positive rates were shown for different depths in simulated count datasets. Poisson and NB distributions were used for simulating technical and biological replicate data, respectively.